# ENHANCED OPTICAL COOLING OF PARTICLE BEAMS IN STORAGE RINGS*

E.G.Bessonov[#], Lebedev Phys. Inst. RAS, Moscow

*Abstract*

A method of enhanced optical cooling (EOC) based on nonlinear selective interaction between particles and theirs amplified undulator radiation wavelets (URW) in storage rings is discussed. It leads to non-exponential fast damping. The selectivity is arranged by a moving screen located on the image plane of the optical system projecting URW there.

## INTRODUCTION

Beam cooling is the reduction of its normalized six dimensional (6D) emittance (hypervolume). Multicycle injection and production of very dense particle beams in storage rings is possible if 6D cooling is used.

The 6D emittance and the sums of 4D and 2D beam emittances, according to the Liouville's theorem are invariants for conservative Hamiltonian system if the intrabeam interactions are neglected and if the density is continuous function of particle position and momentum. External fields could be arbitrary (linear, nonlinear).

In reality the system have a friction, the intrabeam interactions occur, the number of particles and the distance between particles are finite (countable set of particles). It means that the theorem does not work if we: 1) Introduce in the system a friction to convert it into non-Hamiltonian one. Friction in storage rings, in case of linear systems, leads to exponential decay. There is a correlation of damping decrements (positive or negative) determined by Robinson's damping criterion [1], [2], [3] which can limit the rate of particle cooling. 2) Inelastic interactions of particles in the beam with excitation of electronic or nuclear transitions and following emission of photons can lead to cooling. 3) Because of countable set of particles, the interparticle space can be populated by extraneous particles if external forces act over distances smaller than interparticle spacing both in ordinary 3D space or in 3D momentum space or in 6D phase space. Fast microwave electronics, thin and short laser-like URWs and monochromatic laser light exciting electronic or nuclear transitions of ions can be used in the last case. Stochastic cooling, optical stochastic cooling and considered below method of EOC are here.

## ENHANCED OPTICAL COOLING

The scheme of EOC includes pick-up (PU) and kicker (KU) undulators, lattice with zero local slippage factor $\eta_{c,l}=0$ and phase advance between undulators $\psi_x^{bet} \cong 2\pi(k+1/2)$, $k=1,2,3...$, optical system consisting of optical filter, lenses, optical line with variable time delay, movable screen located in the image plane and optical amplifier (see Fig.1) [4]-[8]. URW's emitted in the PU are focused by a lens on the image plane. Movable screen open URWs emitted by particles with maximal positive deviations of their position from synchronous one. They are amplified. The initial phase of particles in the fields of their URWs in KU is chosen so that the rate of the energy loss of the particles is maximal. It does not depend on the deviation of the particle energy from synchronous one. At the same time the rate is zero for particles with negative deviations from synchronous one as the screen never open their URWs in the image plane.

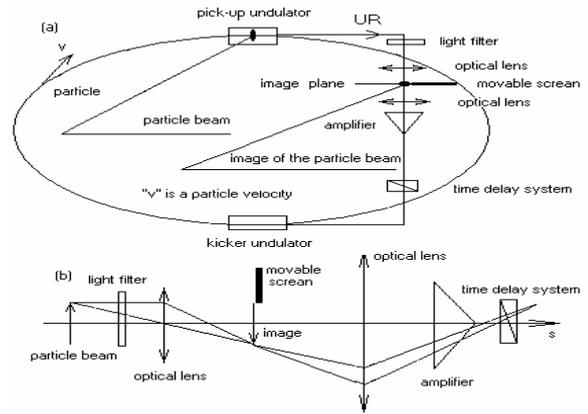

*Fig.1. The scheme of the EOC of particle beams in a storage rings (a) and unwrapped optical scheme (b).*

In the EOC scheme every particle enters the kicker undulator together with their URW in decelerating phase and looses energy. The change of the amplitude of the particle betatron oscillations is $\delta A_x^2 = -2x_{\beta,k,0}\delta x_\eta + (\delta x_\eta)^2 \simeq -2x_{\beta,k,0}\delta x_\eta$, where $x_{\beta,k,0}$ is the initial particle deviation from it's closed orbit in kicker undulator; $\delta x_\eta = \eta_x \beta^{-2}(\delta E/E)$ is the jump of it's closed orbit; $\eta_x$ is the dispersion function of the storage ring; $\beta$ is the normalized velocity. In the case $x_{\beta,k} < 0$, $\delta x_\eta < 0$ the damping both in longitudinal and transverse directions takes place (see Fig.2).

Using short (the length $\sim M\lambda_{1,min}$) and narrow (the diameter $\sim 2\sigma_w$) laser-like wavelets permits to give a chance to control the behavior of individual particles in the storage rings without significant disturbing another ones. Here $M$ is the number of the undulator periods, $\lambda_{1,min} = \lambda_u(1+K^2)/2\gamma^2$ is the minimum UR wavelength, $\lambda_u$ is the undulator period, $K = Ze\sqrt{\overline{B^2}}\lambda_u/2\pi M_p c^2$ is the deflection parameter, $\overline{B^2}$ is the average of the square of the undulator magnetic field strength, $Ze$ and $M_p$ are the particle charge and mass.

_______________
*Supported by RFBR under grant No 05-02-17162.
[#]Corresponding author; bessonov@x4u.lebedev.ru

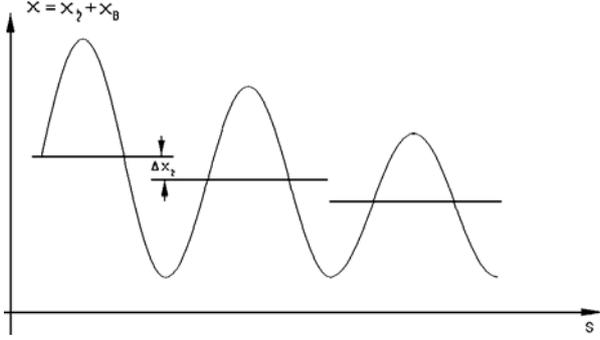

*Fig.2. Motion of a particle in longitudinal-radial plane. The deviation of the particle from synchronous orbit is $x = x_\eta + x_b$, where $x_\eta$ is the deviation of the closed orbit of the particle from synchronous one and $x_b$ is the deviation of the particle from the closed orbit, $\Delta x_\eta$ is jumps of the closed orbit.*

Three schemes of EOC of particle can be realized in general case using 2 or more kicker undulators and a lattice with zero local slippage factors between them.

In the first scheme undulators are installed in straight sections at a distance determined by a betatron phase advance for the lattice segment $(2p+1)\pi$ between pick-up and first kicker undulator and $2p'\pi$ between next kicker undulators; where $p, p' = $ 1,2,3... are integer numbers. In this case the energy loosed by particles is accompanied by a decrease in both energy spread and amplitudes of betatron oscillations of the beam. So the EOC is going both in the longitudinal and transverse degrees of freedom.

Second modification of EOC method uses the pick-up undulator followed by even number of kicker undulators installed in straight sections of a storage ring at distances determined by the phase advance $(2p+1)\pi$. In this scheme, the deviations of particles in undulators "$i$" and "$i+1$" are $x_{\beta_i} = -x_{\beta_{i+1}}$ and that is why the decrease of energy of particles in undulators does not lead to change of the particle's betatron amplitude at the exit of the last undulator. So it leads the cooling of the particle beam in the longitudinal plane only.

A third modification of EOC method uses the scheme of the second one. However, if the optical delay line is tuned such a way that particle decreases its energy in odd undulators and increases it in even ones, then the change of the energy of the particle in undulators leads to decrease of their betatron amplitudes and do not lead to change of their energy at the exit of the last undulator. In this case cooling of the particle beam is going in the transverse plane only.

The URW emitted by a particle in the PU undulator and amplified interact efficiently with the same particle in the KU. Radiation from one particle does not disturb trajectories of other particles if an average distance between particles in a longitudinal direction is more, than the length of the URW, $M\lambda_{UR}$, where $M$ is the number of the undulator periods; $\lambda_{UR}$ is the wavelength of the emitted undulator radiation (UR). This case is named "single particle in the sample". It corresponds to the beam current

$$i < i_c = \frac{Zec}{M\lambda_{UR}} = \frac{4.8 \cdot 10^{-9} Z}{M\lambda_{UR}}[A] \qquad (1)$$

If overlapping of other particles with URW occurs, then URWs does not disturb the amplitudes of betatron and synchrotron oscillations of other beam particles in the first approximation and leads to their increase of in the second one because of a stochasticity of the initial phase of the URWs for other particles.

The total energy radiated by a relativistic particle traversing a given undulator magnetic field $B$ of finite length is given by

$$E_{tot} = \tfrac{2}{3} r_p^2 \overline{B^2} \gamma^2 M\lambda_u, \qquad (2)$$

where $\overline{B^2}$ is an average square of magnetic field strength along the undulator length $L_u = M\lambda_u$; $\gamma$ is the relativistic factor; $r_p = Z^2 e^2/M_p c^2$ is the classical radius of the particle.

The number of photons in the URW emitted by electrons in suitable for cooling frequency and angular ranges $(\Delta\omega/\omega)_c = 1/M$, $(\Delta\vartheta)_c = \sqrt{(1+K^2)/M}$ is defined by

$$N_{ph} = \frac{\Delta E_1^{cl}}{\hbar\omega_{1\max}} \approx \pi\alpha \frac{K^2}{1+K^2}, \qquad (3)$$

where $\Delta E_1^{cl} = (dE_1^{cl}/d\omega)\Delta\omega = 3E_{tot}/2M(1+K^2)^2$, $\omega_{1\max} = 2\pi c/\lambda_{1\min}$, $\alpha = e^2/\hbar c \cong 1/137$ [8].

If the density of energy in the URWs is approximated by Gaussian distribution, the R.M.S. electric field strength $E_w^{cl}$ of the wavelet of the length $2M\lambda_{1\min}$ and the diameter $2\sigma_w > 2\sigma_{w,c}$ in the kicker undulator is defined by the expression

$$E_w^{cl} = \sqrt{\frac{\Delta E_1^{cl}}{M\sigma_w^2 \lambda_{1\min}}} = \frac{\sqrt{2}r_p\gamma^2\sqrt{\overline{B^2}}}{(1+K^2)^{3/2}\sqrt{M}\sigma_w}. \qquad (4)$$

where the waist size $\sigma_{w,c} = \sqrt{L_u\lambda_{1\min}/8\pi}$ corresponds to the Rayleigh length $Z_R = L_u/2$, $Z_R = 4\pi\sigma_{w,c}^2/\lambda_{1\min}$.

The electric field value (4) is valid for $N_{ph} \gg 1$. Such case can be realized only for heavy ions with atomic number $Z > 10$, $K > 1$. If $N_{ph} < 1$ then, according to quantum theory, one photon is emitted with the probability $p_{em} = N_{ph}$, and the energy $\Delta E_1 = \hbar\omega_{1,\max}$. In this case the maximum rate of energy losses for the electron in the fields of the kicker undulator and amplified URW is

$$P_{loss}^{\max} = -eE_w^{cl} L_u \beta_{\perp m} f \Phi(N_{ph}) N_{kick} \sqrt{\alpha_{ampl}}, \qquad (5)$$

where $\beta_\perp = K/\gamma$; $f$ is the revolution frequency; $N_{kick}$ is the number of kicker undulators; and $\alpha_{ampl}$ is the gain in the optical amplifier. The function $\Phi(N_{ph})|_{N_{ph}\gg 1} = 1$, $\Phi(N_{ph})|_{N_{ph}\ll 1} = \sqrt{N_{ph}}$.

The damping times for the longitudinal and transverse degrees of freedom are

$$\tau_{s,EOC} = \frac{6\sigma_{E,0}}{|P_{loss}^{\max}|}, \quad \tau_{x,EOC} = \tau_{s,EOC}\frac{\sigma_{x,0}}{\sigma_{x_\eta,0}} = \frac{6\beta^2 E_s \sigma_{x,0}}{|P_{loss}^{\max}||\eta_{x,\,kick}|}, \quad (6)$$

where $\sigma_{E,0}$ is the initial energy spread of the electron beam, $\sigma_{x,0}$, $\sigma_{x_\eta,0}$ are the initial radial beam dimensions determined by betatron and synchrotron oscillations, $\eta_{x,\,kick} \neq 0$ is the dispersion function in the kicker undulator. The damping time for the longitudinal direction does not depend on $\eta_{x,\,kick}$ and for the transverse one is inverse to $\eta_{x,\,kick}$. Factor 6 takes into account that the initial energy spread is $2\sigma_{E,0}$, electrons does not interact with their URWs every turn, the jumps of the electron closed orbit lead to lesser jumps of the amplitude of synchrotron and betatron oscillations [6].

According to (6), the damping time in the longitudinal plane is short (proportional to the energy spread, not to the initial energy of particles). Moreover, because of non-exponential decay the degree of cooling can be much higher than 1/e reduction of the beam emittance.

The average power of the optical amplifier is

$$P_{ampl} = \varepsilon_{sample} \cdot f \cdot N_e + P_n, \quad (7)$$

where $\varepsilon_{sample} = \hbar\omega_{1,\max} N_{ph} \alpha_{ampl}$ is the average energy in an URW, $N_e$ stands for the number of electrons in the ring, $P_n$ is the noise power.

If $\eta_{c,l} \neq 0$, the difference in the propagation time of the URW $dt$ and the traveling time $T_{p,k}$ of the electron between pickup and kicker undulators depends on initial conditions of electron's trajectory which can be expressed as

$$cdt = c_t - R_{51}(s,s_0)\cdot x_0 - R_{52}(s,s_0)\cdot x'_0 - R_{56}(s,s_0)\frac{\Delta E}{\beta^2 E}|_{R_{51}=R_{52}=0} \cong c_t + cT_{p,k}\eta_{c,l}\frac{\Delta E}{\beta^2 E}. \quad (8)$$

where $c_t$ is a constant.

The initial phase of an electron in the field of amplified URW propagating through kicker undulator is $\varphi_{in} = \omega_{1,\max} dt$ and the rate of the energy loss

$$P_{loss} = -|P_{loss}^{\max}|\sin(\varphi_{in})\cdot f(\Delta E), \quad (9)$$

where $f(\Delta E) = 1 - |\varphi_{in}(\Delta E)|/2\pi M$, if $|\varphi_{in}| \leq 2\pi M$ and $f(\Delta E) = 0$ if $|\varphi_{in}| > 2\pi M$, $\Delta E = E - E_d$. The function $f(\Delta E)$ takes into account that electron with some energy $E_d$ and its URW enter kicker undulator at the phase $\varphi_{in} = 0$ and passing together all undulator length at zero rate of the energy loss if $c_t = 0$.

According to (9), electrons with different initial phases are accelerated or decelerated and gathered at phases $\varphi_{in}^m = \pi + 2\pi m$ ($-M \leq m \leq M$, $m = 0, \pm 1, ..\pm M$) and at energies

$$E_m = E_d + \frac{(2m+1)\pi\beta^2}{\omega_{1,\max}T_{p,k}\,\eta_{c,l}}E_d, \quad (10)$$

if RF accelerating system is switched off (see Fig.2).

The energy gaps between equilibrium energy positions have magnitudes given by

$$\delta E_{gap} = E_{m+1} - E_m = \frac{\lambda_{1,\min}}{L_{p,k}\eta_{c,l}}\beta^2 E_d, \quad (11)$$

where $L_{p,k} = cT_{p,k}$.

The power loss $P_{loss}$ is the oscillatory function of the energy $|E - E_d|$ with the amplitude linearly decreasing from the maximum value $|P_{loss}^{\max}|$ at the energy $E = E_d$ to a zero one at the energy $|E - E_d| \geq M\cdot\delta E_{gap}$. If the particle energy falls into the energy range $|E - E_d| < M\cdot\delta E_{gap}$ it is drifting to the nearest energy value $E_m$. The variation of the particle's energy looks like it produces aperiodic motion in one of $2M$ potential wells located one by one. The depth of the well decreases with their number $|m|$. If the delay time in the optical line is changed, the energies $E_m$ are changed as well.

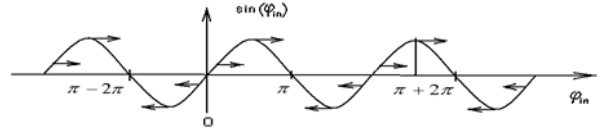

Fig. 3: In the EOC scheme electrons are grouping near the phases $\varphi_{in} = \pi + 2\pi m$ (energies $E_m$).

The space resolution of the optical system is

$$\delta x_{res} \cong 1.22\lambda_{1\min}/(\Delta\theta)_c = 0.86\sqrt{\lambda_{1\min}L_u}. \quad (12)$$

Damping time of lead ions in LHC at the energy $\gamma = 953$ can be ~5 sec if $\alpha_{ampl} \approx 10^8$, $\lambda_{1\min} = 5.5\cdot 10^{-5}$ cm, $P_{ampl} \approx 250$ W [6].